\documentclass[letterpaper, 12pt, reqno, fleqn]{article}
\usepackage[top = 1in, bottom = 1in, left = 1in, right = 1in]{geometry}
\usepackage{lmodern}
\usepackage[T1]{fontenc} 
\usepackage{natbib}
\usepackage{url}
\usepackage[colorlinks,citecolor=blue,linkcolor=blue,urlcolor=blue,pagebackref]{hyperref}
\usepackage[classfont = sanserif, langfont = roman, funcfont = italic]{complexity}
\usepackage{amsmath, amssymb, amsthm}
\usepackage{algorithm}
\usepackage{algpseudocode}
\usepackage[dvipsnames]{xcolor}
\usepackage{authblk}
\usepackage{cleveref}
\usepackage{orcidlink}
\usepackage{dsfont}

\newtheorem{theorem}{Theorem}[section]
\newtheorem{lemma}[theorem]{Lemma}
\newtheorem{claim}[theorem]{Claim}
\newtheorem{corollary}[theorem]{Corollary}
\newtheorem{conjecture}[theorem]{Conjecture}
\newtheorem{definition}[theorem]{Definition}

\theoremstyle{remark}
\newtheorem{remark}[theorem]{Remark}

\newclass{\SETH}{SETH}
\newclass{\NUSETH}{NUSETH}
\newclass{\QSETH}{QSETH}
\newclass{\NUQSETH}{NUQSETH}

\newlang{\FOV}{\ensuremath{\mathcal{F}\mathrm{OV}}}
\newlang{\OV}{\ensuremath{\mathrm{OV}}}
\newlang{\PARITY}{PARITY}
\newcommand{\kOV}{\ensuremath{k\text{-}\mathrm{OV}}}
\newcommand{\softO}{\widetilde{O}}
\newcommand{\pexp}[1]{p_{#1}}
\newcommand{\dval}{\left \lceil \log^2 n \right \rceil}

\providecommand{\keywords}[1]
{
  \small	
  \textbf{Keywords:} #1
}

\begin{document}

\title{Uniform Quantum Average-Case Fine-Grained Hardness of Arithmetized $k$-Orthogonal Vectors}

\author{Dhruv Prasad Deshpande\orcidlink{0009-0005-1141-5169}\footnote{Department of Computer Science and Information Systems, Birla Institute of Technology and Science, Pilani, Pilani-333031, Rajasthan, I\textsc{ndia}. Email: \texttt{f20230588@pilani.bits-pilani.ac.in}.} \and Abhishek Mishra\orcidlink{0000-0002-2205-0514}\footnote{Department of Computer Science and Information Systems, Birla Institute of Technology and Science, Pilani, Pilani-333031, Rajasthan, I\textsc{ndia}. Email: \texttt{abhishek.mishra@pilani.bits-pilani.ac.in}.}}

\maketitle
\thispagestyle{empty}

\maketitle

\begin{abstract}

We study exact evaluation of the arithmetized $k$-Orthogonal Vectors polynomial $\FOV^k$ of \cite{BRSV17, BRSV18}. For every fixed integer $k \geq 3$, set $d(n) = \dval$. Assuming the basic decision version of \textit{Quantum Strong Exponential-Time Hypothesis} ($\QSETH$), for every constant $\varepsilon > 0$, no uniform quantum algorithm runs in $O(n^{k/2 - \varepsilon})$ time on every input and evaluates this polynomial correctly with probability at least $3 / 4$ over a uniformly random input in $\mathbb F_{p(n)}^{knd(n)}$ and the algorithm's internal randomness, for all sufficiently large $n$. This holds when $p(n)$ is the smallest prime greater than $n^{k - 1}$ and when it is the smallest prime greater than $n^k$. Our direct reduction combines classical polynomial self-correction with calls to a quantum average-case evaluator. For the smaller prime, random sampling finds an orthogonal tuple with constant probability whenever the count is large enough to wrap around modulo $p(n)$; otherwise, the exact residue determines the answer. For the larger prime, quantum minimum finding and an asymptotic prime-gap bound account for the cost of generating the prescribed field. In view of prior isolation-based reductions permitting small characteristics, we present these results as explicit reductions for the two canonical-prime families, with all setup costs included in
the uniform quantum running time. The hardness guarantee concerns single-instance exact evaluation; a matching quantum evaluation algorithm and a quantum proof-of-work protocol remain open.


\end{abstract}

\newpage

\pagenumbering{roman}
\setcounter{page}{1}

\maketitle

\tableofcontents

\newpage

\pagenumbering{arabic}
\setcounter{page}{1}

\section{Introduction}
\label{sec:intro}

Assuming the \textit{Quantum Strong Exponential-Time Hypothesis} ($\QSETH$) \citep{ACLWZ20}, for every fixed integer $k\geq 3$, we show that $k\OV$ requires $\Omega \left( n^{k/2-o(1)} \right)$ time for uniform quantum algorithms. This matches the running time of the straightforward quantum algorithm obtained by applying Grover's algorithm\citep{Gro96} to the $n^k$ candidate $k$-tuples, and therefore gives a conditional optimality result for quantum $k\OV$. We also state the corresponding non-uniform result under $\NUQSETH$ ($\QSETH$ hypothesized for non-uniform algorithms with $O \left( 2^{(1-\epsilon)n / 2} \right)$ bits of advice). These reductions are direct, and are stated for completeness.

We then study the effect of this hardness assumption on the $\mathcal{F}\mathrm{OV}^k$ family of polynomials as proposed  by \cite{BRSV17,BRSV18}. The non-uniform quantum hardness of $\mathcal{F}\mathrm{OV}^k$ follows from the existing worst-case to average-case reduction. The reduction uses the same random self-reducibility argument as in the classical setting, and the prime defining the finite field can be supplied as non-uniform advice. Thus $\NUQSETH$ gives the corresponding $\Omega \left( n^{k/2 - o(1)} \right)$ average-case lower bound for non-uniform quantum algorithms. The uniform setting is different because all parameters depending on the input length must be generated by the reduction itself. In particular, the prime defining the field can no longer be supplied as advice.

For the canonical prime families studied here, the reduction must include the cost of generating the prescribed prime. The direct counting argument with $p>n^k$ avoids modular wrap-around. Using the classical prime-generation bound in Lemma~\ref{lem:prime} at this scale gives an upper bound that does not supply a fixed polynomial saving below $n^{k / 2}$. This is a limitation of that particular parameter choice and construction, not a general obstruction to uniform quantum average-case hardness; we discuss prior isolation and extension-field approaches below. Our first reduction instead uses the smallest prime greater than $n^{k - 1}$.
If the number of orthogonal $k$-tuples is at most $n^{k - 1}$, there is no wrap-around. If it exceeds $n^{k - 1}$, a uniformly sampled tuple is orthogonal with probability greater than $1/n$, so $2n$ independent samples find a witness with probability greater than $1 - e^{-2}$.
Combining these cases with polynomial self-correction yields the $\Omega(n^{k/2 - o(1)})$ average-case lower bound for this canonical family against uniform quantum algorithms.

We also recover the original choice of a prime larger than $n^k$, although the resulting statement is only asymptotic. We use an unconditional bound on prime gaps \citep{BHP01} together with quantum minimum finding \citep{DH96} to search for a suitable prime in the guaranteed interval. The cost of this search is below the $n^{k/2-o(1)}$ scale for sufficiently large $n$. The larger-prime construction uses an asymptotic prime-gap theorem. We do not derive explicit numerical thresholds or concrete implementation bounds. The result establishes the stated asymptotic reduction for this canonical field choice.

\subsection{Our Contributions}
\label{sec:our-contributions}

Our first contribution is a quantum fine-grained hardness result for $k\OV$. Assuming $\QSETH$, for every fixed integer $k\geq 3$, we establish an $\Omega \left( n^{k/2-o(1)} \right)$ lower bound for uniform quantum algorithms for $k\OV$. The proof is obtained by reducing bounded-width $\SAT$ to $k\OV$ and is the quantum analogue of the standard $\SETH$-based hardness result. We also give the corresponding statement for non-uniform quantum algorithms under $\NUQSETH$.

Our main contribution is a direct sampling-based worst-case to average-case reduction for $\FOV^k$ with a prescribed canonical prime, including its generation cost in the uniform quantum time bound. For $k \geq 3$, we show that if $k\OV$ requires $\Omega \left( n^{k/2 - o(1)} \right)$ time for uniform quantum algorithms, then $\FOV^k$ requires $\Omega \left( n^{k/2 - o(1)} \right)$ time on average when the underlying prime is the smallest prime satisfying $p > n^{k - 1}$. We further show that the larger modulus $p > n^k$ (smallest prime greater than $n^k$) used in the original construction also gives the same asymptotic quantum lower bound for uniform algorithms, after accounting for the cost of finding the prime.

\section{Related Work and Scope}
\label{sec:related}

\cite{BRSV17}, Appendix~E of the full version, already give an isolation reduction from two-list $\OV$ to unique-$\OV$, adding $4 \log n + 2$ coordinates with only polylogarithmic overhead. They explain that uniqueness removes the need for the characteristic to exceed the maximum witness count; a characteristic on the order of the dimension suffices for their polynomial self-correction. Our sampling argument handles $k$ lists directly and retains the original Boolean instance for the polynomial call, avoiding the isolation transformation and its additional coordinates. Thus, small-characteristic arithmetization already has a precedent.
The contribution here is an explicit alternative reduction for the specified canonical-prime families, with the quantum time budget made explicit. In addition, \cite{BRSV18}, Remarks~2.2 and~2.6, already discuss uniform setup and extension fields. Uniformity by itself is therefore not a new feature of the general arithmetization approach.

Quantum worst-case to average-case reductions for linear problems provide a different comparison. \cite{AGGSS24} give uniform quantum reductions for arbitrary fixed linear maps, including over small fields, from average success $\alpha>0$ that may decrease with the input length. Writing $m$ for their vector dimension and $T$ for the average-case algorithm's cost, their bound is $\widetilde O((T + m^{3 / 2})\operatorname{poly}(1 / \alpha))$ in their coherent matrix-access model, suppressing logarithmic field-size factors. \cite{AK25} treat matrix-vector multiplication with both the matrix and vector uniformly distributed. Their explicit quantum oracle reduction uses $O(\alpha^{-2})$ calls to the average-case algorithm and $\widetilde O((Q(m) + m^{3 / 2})\alpha^{-2})$ matrix-oracle queries, where $Q(m)$ is the original query bound. This is a query-complexity statement; it must be distinguished from a bound that charges for constructing or accessing an explicitly stored matrix. \cite{AHS26} obtain a uniform reduction in time $\widetilde O(mT^{1 / 3})$ for $m \leq T \leq m^{3 / 2}$, assuming constant field size and constant positive average success, with coherent access
to the average-case oracle and matrix entries. These coherent-access requirements are part of the respective models.

Our average-case guarantee is the fixed threshold $3 / 4$ for uniformly random field inputs. The polynomial corrector uses fresh executions of the evaluator on classically specified inputs and measures its field-valued outputs; it does not require coherent access to that evaluator or its inverse. For our parameters its cost is $\widetilde O(n + t(n))$, in addition to prime generation. The reduction for $p_k$ separately uses coherent primality testing inside quantum minimum finding. The linear reductions cited above exploit linearity of the target map. Here the map from the input vector coordinates to the output residue is a nonlinear polynomial of total degree $kd$, so applying those reductions would require a further argument preserving the dimension, input distribution, and running-time bounds. Our results establish single-instance evaluation hardness; they do not establish a quantum proof-of-work protocol or hardness at arbitrary small success probabilities.

\section{Preliminaries}
\label{sec:prelims}

Throughout, $[n] = \{\, 1, \ldots, n \,\}$, and $\mathbb{F}_p$ denotes the finite field
over a prime $p$. All logarithms are base $2$ unless stated otherwise. $r\SAT$ denotes a $\SAT$ instance of clause-width $r$. All algorithms are in the Word QRAM model of computation with $O(\log n)$-bit words. 

Throughout, an informal lower bound of the form $\Omega(n^{c-o(1)})$ abbreviates the exclusion, for every constant $\varepsilon>0$, of an $O(n^{c-\varepsilon})$-time algorithm with the stated
success guarantee. For average-case computation, that guarantee is required for all sufficiently large $n$. This shorthand does not assert a pointwise lower bound at every sufficiently large input length; the formal statements below specify the quantifiers explicitly.

\subsection{$\SETH$ and related Hardness Assumptions}

Intuitively, $\SETH$ conjectures that every fixed improvement in the exponential running-time exponent is ruled out for some clause width: for every constant $\varepsilon>0$, there is an integer $r\geq 3$ for which no uniform algorithm solves $r\SAT$ in time $O(2^{(1-\varepsilon)n})$. We differentiate between the uniform and non-uniform versions of $\SETH$ for clarity.

\begin{conjecture}[Strong Exponential-Time Hypothesis ($\SETH$) \citep{Impagliazzo2001b}]
\label{col:SETH}

For every $\varepsilon > 0$, there exists an integer $r \geq 3$ such that no uniform algorithm can solve $r\SAT$ in time
$O \left( 2^{(1-\varepsilon)n} \right)$.

\end{conjecture}

\begin{conjecture}[Non-Uniform $\SETH$ ($\NUSETH$) \citep{Stephens2019, GKASS26}]
\label{col:NUSETH}

For every $\varepsilon > 0$, there exists an integer $r \geq 3$ such that no non-uniform algorithm can solve $r\SAT$ in time $O \left( 2^{(1-\varepsilon)n} \right)$, using $O \left( 2^{(1 - \varepsilon)n} \right)$ bits of advice.

\end{conjecture}

For hardness against non-uniform algorithms, the relevant assumption is the non-uniform $k\OV$ conjecture, which is implied by $\NUSETH$; this distinction is explicitly discussed in~\cite[Appendix~A.1]{MJPAG24}.

\begin{conjecture}[Quantum Strong Exponential-Time Hypothesis ($\QSETH$) \citep{ACLWZ20}]
\label{assum:qseth}

For every $\varepsilon > 0$, there exists an integer $r \geq 3$ such that no uniform bounded-error quantum algorithm can solve $r\SAT$ in time $O \left( 2^{(1-\varepsilon)n/2} \right)$.

\end{conjecture}

\begin{conjecture}[Non-Uniform Quantum Strong Exponential-Time Hypothesis ($\NUQSETH$)]
\label{col:NUQSETH}

For every $\varepsilon>0$, there exists an integer $r\geq 3$ such that no non-uniform bounded-error quantum algorithm solves $r\SAT$ in time $O(2^{(1-\varepsilon)n/2})$ with classical advice of length $O(2^{(1 - \varepsilon) n / 2})$.

\end{conjecture}

\subsection{Orthogonal Vectors and Fine-Grained Hardness Assumptions}

Given two sets of $n$ vectors, the OV problem asks whether there exists a pair of vectors (one from each set) such that they are orthogonal to each other.

\begin{definition}[Orthogonal Vectors ($\OV$)]
\label{def:ov}

The $\OV$ problem on vectors of dimension $d$ is to determine, given two sets $U_1$ and $U_2$ of $n$ vectors from $\{\, 0, 1 \,\}^{d(n)}$ each, whether there exist $u_1 \in U_1$ and $u_2 \in U_2$ such that, over the integers,
\[
\sum_{\ell \in [d(n)]} u_{1\ell}u_{2\ell} = 0.
\]

\end{definition}

The $k$-Orthogonal Vectors ($k\OV$) problem is a generalization of the Orthogonal Vectors ($\OV$) problem over $k$ such sets.

\begin{definition}[$k$-Orthogonal Vectors \citep{BRSV18}]
\label{def:kov}

Fix an integer $k \ge 2$. An instance of $k\OV$ consists of $k$ sets $U_1,\ldots,U_k$, each containing $n$ Boolean vectors of dimension $d(n)$. The task is to decide whether one can
select a vector $u_s$ from each set $U_s$ so that
\[
    \sum_{\ell=1}^{d(n)} \prod_{s=1}^{k} u_{s\ell} = 0,
\]
where the arithmetic is performed over the integers.
Throughout, we take $d(n)=\dval$ unless specified otherwise.

\end{definition}

The classical hardness of $k\OV$ is known to follow from $\SETH$ \citep{Wil05}, and
is the standard basis for a large family of fine-grained lower bounds.

\begin{theorem}
[Uniform hardness of $k\OV$ \citep{Wil05}]
\label{theorem:kOV}

Assuming $\SETH$, for every fixed integer $k\geq 2$ and every constant $\varepsilon>0$, no uniform algorithm solves $k\OV$ in $O(n^{k-\varepsilon})$ time in the worst case.

\end{theorem}

\begin{theorem}[Non-uniform hardness of $k\OV$ \citep{Wil05}]

Assuming $\NUSETH$, for every fixed integer $k\geq 2$ and every constant $\varepsilon>0$, no non-uniform algorithm solves $k\OV$ in $O(n^{k-\varepsilon})$ time in the worst case using
$O(n^{k-\varepsilon})$ bits of advice.

\end{theorem}

\begin{theorem}
[Uniform quantum hardness of $k\OV$]
\label{thm:uqkov}

Assuming $\QSETH$, for every fixed integer $k\geq 3$ and every constant $\varepsilon>0$, no uniform bounded-error quantum algorithm decides $k\OV$ in $O(n^{k/2-\varepsilon})$ time in the worst case.

\end{theorem}

\begin{theorem}[Non-uniform quantum hardness of $k\OV$]
\label{thm:nuqkov}

Assuming $\NUQSETH$, for every fixed integer $k\geq 3$ and every constant $\varepsilon>0$, no non-uniform bounded-error quantum algorithm decides $k\OV$ in $O(n^{k/2-\varepsilon})$ time in the worst case using $O(n^{k/2-\varepsilon})$ classical bits of advice.

\end{theorem}

Theorems \ref{thm:uqkov} and \ref{thm:nuqkov} can be proven easily by adapting the classical SETH to OV reduction to the quantum setting. The proofs for both these theorems can be accessed in Appendix \ref{app:A}.

\subsection{Prime Fields}

We will be using the following notation to denote various prime computing functions.

\begin{definition}[$\pexp{m}$]

$\pexp{m}(n)$ is the smallest prime larger than $n^m$.

\end{definition}

\begin{lemma}[Implied by \cite{LO87}]
\label{lem:prime}

$\pexp{m}(n)$ can be computed deterministically in $\softO \left( n^{m/2 + \alpha} \right)$ time for any $\alpha > 0$.

\end{lemma}

\subsection{Average-Case Hardness and Random Self-Reducibility}

\begin{definition}[Average-case hardness \citep{BRSV17}]

Let $\mathcal{F}=\{f_n\}$ be a family of functions, and let $D_n$ denote the domain of $f_n$. We say that $\mathcal{F}$ is computable on average in time $t$ under the uniform input distribution if there exists an algorithm $A$ whose running time on every input in $D_n$ is at most $t(n)$ and such that, for all sufficiently large $n$,
\[
    \Pr_{\Pr_{x \sim \mathrm{Unif}(D_n),\,A}}
    \bigl[A(x)=f_n(x)\bigr] \ge \frac{3}{4}.
\]
When $A$ is randomized or quantum, this probability is taken jointly over the uniformly distributed input and the internal randomness of $A$, including its measurement outcomes. For a non-uniform algorithm, its advice depends only on the input length and is fixed before the input is sampled.
\end{definition}

The central tool for establishing average-case hardness of this kind is the classical random self-reducibility of low-degree polynomial evaluation, which we restate here (with running time made explicit, following \cite{BRSV18}).

\begin{lemma}[Random self-reducibility of low-degree polynomials \citep{Lip91,FF90,GS92,BRSV17,BRSV18}]
\label{lem:selfred}

Let $N,D$ be positive integers with $D>9$, let $p>12D$ be prime, and fix $\epsilon\in(0,1/3)$. Suppose that $f:\mathbb{F}_p^N\to\mathbb{F}_p$ is a polynomial of degree $D$ and that an algorithm $A$, running in time $t$, satisfies
\[
    \Pr_{x\sim\mathrm{Unif}(\mathbb{F}_p^N), A}
    \bigl[A(x)=f(x)\bigr]\ge 1-\epsilon.
\]
Then there exists a randomized algorithm $B$ with running time
\[
    O\!\left(ND^2\log^2 p+D^3+tD\right)
\]
that, on every input $x\in\mathbb{F}_p^N$, outputs $f(x)$ with probability at least $2/3$:
\[
    \forall x\in\mathbb{F}_p^N,\qquad
    \Pr_B\bigl[B(x)=f(x)\bigr]\ge \frac{2}{3}.
\]
Thus, correctness on a sufficiently large fraction of uniformly distributed inputs can be converted into bounded-error correctness on every input, with the stated running-time overhead.

\end{lemma}

\begin{remark}[Lemma~\ref{lem:selfred} is true for quantum algorithms $A$ and $B$ as well]
\label{rem:agnostic}

The algorithm $B$ constructed in the proof of Lemma~\ref{lem:selfred} consists of
classical procedures and \emph{black-box} calls to $A$. Nothing about this construction relies on $A$ being classical: if $A$ is a quantum algorithm, $B$ is simply a hybrid algorithm that queries a quantum subroutine, and its running time is exactly as stated, with $t$ now denoting the running time of the quantum algorithm $A$.

\end{remark}

\subsection{The $\FOV^k$ Family}

\begin{definition}[$\FOV^k$ \citep{BRSV18}]
\label{def:fovk}

Fix an integer $k\ge 2$ and a prime $p$. View an input in $\mathbb{F}_p^{knd}$ as $k$ lists $U_1,\ldots,U_k$, each containing $n$ vectors in $\mathbb{F}_p^d$. Write $u_{s,i,\ell}$ for the $\ell$th coordinate of the $i$th vector in $U_s$. We define the polynomial
$f\OV^k_{n,d,p}:\mathbb{F}_p^{knd}\to\mathbb{F}_p$ by
\[
    f\OV^k_{n,d,p}(U_1,\ldots,U_k)
    =
    \sum_{(i_1,\ldots,i_k)\in[n]^k}
    \prod_{\ell=1}^{d}
    \left(1-\prod_{s=1}^{k}u_{s,i_s,\ell}\right),
\]
with all operations performed in $\mathbb{F}_p$. On Boolean inputs, the summand associated with a tuple is $1$ precisely when that tuple is $k$-orthogonal, and is $0$
otherwise. Consequently, the polynomial evaluates to the number of $k$-orthogonal tuples reduced modulo $p$. Its total degree is $kd$.

Choosing dimension $d(n)$ and prime function $p(n)$ determines the family
\[
    \FOV^k_{n, d, p} = \bigl\{f\OV^k_{n,d(n),p(n)}\bigr\}.
\]
We specify these parameter functions separately for the two settings considered below.

\end{definition}

\section{Non-Uniform Quantum Hardness of $\FOV^k$}
\label{sec:nonuniform}

We first record that, for non-uniform algorithms, the classical hardness result of
\cite{BRSV18} for $\FOV^k$ (with $\pexp{\log n}(n)$ the smallest prime exceeding $n^{\log n}$) transfers to the quantum setting without any change to the argument.

\begin{theorem}[Non-uniform quantum hardness of $\FOV^k$]
\label{thm:nonuniform}

Fix an integer $k\geq 3$. Let $p(n)=p_{\log n}(n)$ be the smallest prime greater than $n^{\log n}$, and let $d(n)=\dval$. Suppose that, for every constant $\delta>0$, no non-uniform bounded-error quantum algorithm decides $k\OV$ in $O(n^{k/2-\delta})$ time using $O(n^{k/2-\delta})$ classical bits of advice. Then, for every constant $\varepsilon>0$, there is no non-uniform quantum algorithm $A$ that uses $O(n^{k/2-\varepsilon})$ classical bits of advice, runs in $O(n^{k/2-\varepsilon})$ time on every input, and satisfies
\[
 \Pr_{x\sim\mathrm{Unif}(\mathbb F_{p(n)}^{knd(n)}),\,A}
 \bigl[A(x)=f\OV^k_{n,d(n),p(n)}(x)\bigr]\geq \frac34
\]
for all sufficiently large $n$. The advice is fixed for each input length, and the probability includes the internal randomness of $A$.
 
\end{theorem}

\begin{proof}

We prove the contrapositive. Assume that, for some constant $\epsilon > 0$, a non-uniform quantum algorithm $A$ uses $O(n^{k/2 - \epsilon})$ classical bits of advice, runs in $O(n^{k/2-\epsilon})$ time on every input, and computes $f\OV^k_{n,d,p_{\log n}}$ with average success probability at least $3/4$ for all sufficiently large $n$. We can use the classic random self reduction given by Lemma~\ref{lem:selfred} to get a worst-case algorithm for $f\OV^k_{n,d,p_{\log n}}$. However, in order to do that we need to be able to write down the inputs to $A$, which are composed of elements from $\mathbb{F}_{\pexp{\log n}}$. Therefore we also need to compute $\pexp{\log n}$ in the worst-case algorithm that may require super-polynomial time. Fortunately, we can receive $\pexp{\log n}$ as our non-uniform advice.

Thus, we get an algorithm $B$ that computes $f\OV^k_{n,d,p_{\log n}}$ correctly on \emph{every} input, with probability at least $2 / 3$, in time 

$$O \left( knd (kd)^2 \log^2 p_{\log n} + (kd)^3 + kd O \left( n^{k/2-\epsilon} \right) \right) = \softO \left( n^{k/2-\epsilon} \right)$$ 

(using $d = \dval$ so that the $\log^2 p_{\log n}$ factors, with $p \le \poly(n^{\log n})$, are absorbed into the $\softO(\cdot)$). Since  $k\OV$ trivially reduces to $f\OV^k_{n,d,p_{\log n}}$ ($f\OV^k_{n,d,p}$ returns the number of orthogonal $k$-tuples for a $k\OV$ instance), we get an algorithm that solves $k\OV$ in the worst-case in time $\softO \left( n^{k/2 - \epsilon} \right)$. This contradicts our assumption on the non-uniform quantum hardness of $k\OV$. Thus no such algorithm exists.

\end{proof}

The proof above closely follows the classical argument. We include it to identify the only step that relies on non-uniformity, namely, the availability of $p(n)$ as advice. The self-reduction in Lemma~\ref{lem:selfred} applies equally to classical and quantum
average-case algorithms. Uniform algorithms cannot receive $p(n)$ as advice, so the reduction must compute it within the available running time. We address this obstacle in the next section.

\begin{remark}

This proof does not directly rule out $O(n^{1 - \varepsilon})$-time non-uniform quantum algorithms for $\FOV^2$ on average, for every constant $\varepsilon > 0$, because the reduction overhead is quasi-linear.

\end{remark}

\section{Uniform Hardness of $\FOV^k$}
\label{sec:uniform}

For the canonical prime families studied here, the reduction must include the cost of generating the prescribed prime. The direct counting argument with $p>n^k$ avoids modular wrap-around. Using the classical prime-generation bound in Lemma~\ref{lem:prime} at this scale gives an upper bound that does not supply a fixed polynomial saving below $n^{k / 2}$. This is a limitation of that particular parameter choice and construction,
not a general obstruction to uniform quantum average-case hardness; we discuss prior isolation and extension-field approaches below. Our first reduction instead uses the smallest prime greater than $n^{k - 1}$. If the number of orthogonal $k$-tuples is at most $n^{k - 1}$, there is no wrap-around. If it exceeds $n^{k - 1}$, a uniformly sampled tuple is orthogonal with probability greater than $1 / n$, so $2n$ independent
samples find a witness with probability greater than $1-e^{-2}$. Combining these cases with polynomial self-correction yields the $\Omega(n^{k/2 - o(1)})$ average-case lower bound for this canonical family against uniform quantum algorithms.

\begin{claim}[Uniform hardness of $\FOV^k$]
\label{claim:uniform}

Fix an integer $k\geq 3$. Let $p(n) = p_{k - 1}(n)$ be the smallest prime greater than $n^{k - 1}$, and let $d(n) = \dval$. Suppose that, for every constant $\delta>0$, no uniform randomized algorithm decides $k\OV$ in $O(n^{k - \delta})$ time with error at most
$1/3$ on every input. Then, for every constant $\varepsilon > 0$, there is no uniform algorithm $A$ that runs in $O(n^{k - \varepsilon})$ time on every input and satisfies
\[
 \Pr_{x\sim\mathrm{Unif}(\mathbb F_{p(n)}^{knd(n)}),\,A}
 \bigl[A(x) = f\OV^k_{n,d(n),p(n)}(x)\bigr]\geq \frac34
\]
for all sufficiently large $n$. Here $A$ may be randomized, and the probability includes its internal randomness.

\end{claim}

\subsection{Intuition}

If we choose the smallest prime $p_{k - 1}(n) > n^{k - 1}$, and if we evaluate $\FOV^k_{n,d,p_{k - 1}}$ on an instance of $k\OV$, there are chances that an overflow occurs (since the maximum number of $k$-tuples of orthogonal vectors is $n^k$). In case of an overflow, there is a possibility that $\FOV^k_{n,d,p_{k - 1}}$ evaluates to 0, but the instance of $k\OV$ is a YES instance. This happens when the number of orthogonal $k$-tuples in the $k\OV$ instance is $\alpha p_{k - 1}$, where $\alpha$ is an integer and $1 \le \alpha \le n^k / p_{k - 1}$. In this case $\FOV^k_{n,d,p_{k - 1}}$ evaluates to $\alpha p_{k - 1} \equiv 0 \pmod {p_{k - 1}}$. Therefore, $\FOV^k_{n,d,p_{k - 1}}$ being $0$ does not guarantee (with a high probability, for a given $x$) that no orthogonal $k$-tuple exists in the given $k\OV$ instance. However, since this happens only when the number of orthogonal $k$-tuples is very high, we can simply perform a randomized search to decide $k\OV$ correctly with high probability for such instances.

\subsection{Proof of Claim \ref{claim:uniform}}
\label{proof:uniform}

Assume there exists an algorithm $A$ and $\epsilon > 0$ such that $A$ computes $\FOV^k_{n,d,p_{k - 1}}$ in $O \left( n^{k - \epsilon} \right)$ time on average. The algorithm $B$ that decides $k\OV$ efficiently in the worst-case is defined as follows:

\begin{algorithm}[H]
\caption{Algorithm $B(U_1, \dots, U_k)$}
\begin{algorithmic}[1]
    \For{$2n$ rounds} \Comment{\textbf{Step 1: Randomized Search}}
        \State Uniformly sample $i_1, \dots, i_k \in [n]$
        \If{$u_{1, i_1}, \dots, u_{k, i_k}$ are orthogonal (i.e., $\sum_{\ell=1}^d \prod_{s=1}^k u_{s, i_s, \ell} = 0$)}
            \State \Return \textbf{True}
        \Else
            \State \textbf{continue}
        \EndIf
    \EndFor
    \State Compute the smallest $p_{k-1}(n) > n^{k-1}$ \Comment{\textbf{Step 2: Compute Prime}}
    \State Run classic random self-reduction on $A$ to compute $\mathcal{F}\text{OV}^k_{n,d,p_{k-1}}$ on the worst case. \Comment{\textbf{Step 3: Reduction}}
    \State Let the computed value of $\mathcal{F}\text{OV}^k_{n,d,p_{k-1}}(U_1, \dots, U_k) = c$.
    \If{$c > 0$} \Comment{\textbf{Step 4: Output}}
        \State \Return \textbf{True}
    \Else
        \State \Return \textbf{False}
    \EndIf
\end{algorithmic}
\end{algorithm}

\begin{claim}
Algorithm $B$ decides $k\OV$ efficiently in the worst-case.
\end{claim}

\begin{proof}
Consider two cases for any input $x = (U_1, \ldots, U_k) \in \{\, 0, 1 \,\}^{knd}$\footnote{Since it is a $k\OV$ instance}:

\noindent \textbf{Case 1: Number of orthogonal $k$-tuples in $U_1, \ldots, U_k \le n^{k - 1}$}

Let $\Pr[\text{Algorithm } B \text{ returns True at Step 1}] = \rho_1$.

If the algorithm returns at Step 1, it has a $100\%$ chance of success, as the sampled $i_1, \dots, i_k$ forms a valid $k$-tuple of orthogonal vectors. If the algorithm proceeds to Step 2, classic random self-reduction guarantees a success probability of at least $2 / 3$.

Therefore, the total probability of success is:
\[
\Pr_B[B(U_1, \ldots, U_k) = k\OV(U_1, \ldots, U_k)] \ge \rho_1 \cdot 1 + (1 - \rho_1) \cdot \frac{2}{3} \ge \frac{2}{3}
\]
\noindent \textbf{Case 2: Number of orthogonal $k$-tuples in $U_1, \ldots, U_k > n^{k - 1}$}

Let the number of orthogonal $k$-tuples be $n^{k - 1} + \beta$ for some $\beta$.
In this case, the probability of returning (success) for a single round of step 1 is:
$$\Pr_{i_1,\ldots,i_k \in_R [n]}[u_{i_1}, \ldots, u_{i_k} \text{ are orthogonal}] = \frac{n^{k - 1} + \beta}{n^k} > \frac{1}{n}.$$

Therefore, the probability of an unsuccessful search for a single round of step 1 is:
$$\Pr_{i_1,\ldots,i_k \in_R [n]}[u_{i_1}, \ldots, u_{i_k} \text{ is not orthogonal}] < 1 - \frac{1}{n}.$$

Let $\rho_2$ be the probability of an unsuccessful search over $2n$ rounds of step 1. We have:
\begin{align*}
\rho_2 &= \Pr[\text{unsuccessful search for } 2n \text{ rounds of step 1}] \\
&< \left(1 - \frac{1}{n}\right)^{2n} \\
&< e^{-\frac{2n}{n}} \\
&= e^{-2} \\
&< 0.14
\end{align*}

And the probability that we return in step 1 itself:
$$\Pr[\text{at least one successful search in } 2n \text{ rounds of step 1}] \ge 1 - e^{-2} > 0.86.$$

In this case, classic random self-reducibility makes a zero residue insufficient to decide the Boolean instance (due to wrap-around). Let $\rho_3$ be the probability that step 3 gives the correct result:
$$\rho_3 = \Pr[\text{step 3 gives the correct result}].$$

Therefore,
\begin{align*}
\Pr_B[B(U_1,\ldots,U_k) = k\OV(U_1,\ldots,U_k)] &= (1 - \rho_2) \cdot 1 + \rho_2 \cdot \rho_3 \\
&\ge 0.86 \\
&\ge \frac{2}{3}.
\end{align*}

Since these two cases are mutually exhaustive:
$$\Pr_B[B(x) = k\OV(x)] \ge \frac{2}{3} \quad \text{for any } x \in \{\, 0, 1 \,\}^{knd}.$$

\subsubsection*{Efficiency of B}

It is easy to see that:

\begin{itemize}

\item Step 1 takes $O(nkd) = \softO(n)$ time.
    
\item Using the algorithm from \cite{LO87}, step 2 takes 
$$\tilde{O} \left( n^{\frac{k - 1}{2} + o(1)} \right) = \tilde{O} \left( n^{\frac{k}{2} - \frac{1}{2} + o(1)} \right)$$ 
time.
    
\item Step 3 takes 
$$O \left( knd (kd)^2 \log^2 p_{k - 1} + (kd)^3 + kd O \left( n^{k - \epsilon} \right) \right) = \softO \left( n^{k - \epsilon} \right)$$
time.
    
\item Step 4 takes $O(1)$ time.
    
\end{itemize}

Therefore $B$ runs in $\tilde{O} \left( n^{k - \epsilon} \right)$ time, for some $\epsilon > 0$. Therefore $B$ decides $k\OV$ efficiently on the worst case.

\end{proof}

Since this contradicts the uniform hardness of the $k\OV$ conjecture, no such algorithm $A$ exists. Therefore, for every constant $\varepsilon > 0$, no uniform algorithm computes $\FOV^k_{n,d,p_{k-1}}$ on average in $O(n^{k - \varepsilon})$ time.

\section{Uniform Quantum Hardness of $\FOV^k$}
\label{sec:quniform}

The sampling reduction above also applies when the average-case evaluator is quantum. For $p_{k - 1}$, the classical prime-generation algorithm already fits the required quantum time budget; the construction for $p_k$ below uses quantum minimum finding to obtain the required saving.

\begin{claim}[Uniform quantum hardness of $\FOV^k_{n,d,p_{k - 1}}$]
\label{claim:quniform}

Fix an integer $k \geq 3$. Let $p(n) = p_{k - 1}(n)$ be the smallest prime greater than $n^{k - 1}$, and let $d(n) = \dval$. Suppose that, for every constant $\delta > 0$, no uniform bounded-error quantum algorithm decides $k\OV$ in $O(n^{k/2 - \delta})$ time. Then, for every constant $\varepsilon > 0$, there is no uniform quantum algorithm $A$ that runs in $O(n^{k/2 - \varepsilon})$ time on every input and satisfies
\[
 \Pr_{x\sim\mathrm{Unif}(\mathbb F_{p(n)}^{knd(n)}),\,A}
 \bigl[A(x) = f\OV^k_{n,d(n),p(n)}(x)\bigr] \geq \frac34
\]
for all sufficiently large $n$. The probability is over the uniform input and the internal randomness of $A$.

\end{claim}

\begin{proof}

The proof is identical to the proof for classical uniform algorithms~\ref{proof:uniform}. The only difference is in the efficiency of $A$. Here step 3 runs in $\softO \left( n^{k/2-\epsilon} \right)$ time. With $p = p_{k - 1}(n)$ and $t(n) = O \left( n^{k / 2 - \epsilon} \right)$, the complete decisional algorithm $B$ therefore takes time
$$T_B(n) = \softO \left( n + n^{(k - 1) / 2 + \eta} + n^{k / 2 - \epsilon} \right)$$
for every $\eta > 0$. By taking $\eta  = 1 / 4$, after absorbing the polylogarithms, and taking
$$\delta = \min\{\, \epsilon / 2, 1 / 8 \,\},$$
we get 
$$T_B(n) = O \left( n^{k / 2 - \delta} \right).$$
This contradicts uniform quantum hardness of $k\OV$ and thus, no such algorithm exists.

\end{proof}

We next establish the same form of average-case hardness for $\FOV^k_{n,d,p_k}$ against uniform quantum algorithms. The prime-gap bound used in the reduction applies only for sufficiently large $n$; let $n_0$ denote a threshold beyond which that prime-gap bound applies. This threshold concerns the reduction's prime-generation step, rather than a separate claim of hardness at every individual input length.

\begin{claim}[Uniform quantum hardness of $\FOV^k_{n,d,p_k}$]
\label{claim:quniform2}

Fix an integer $k \geq 3$. Let $p(n) = p_{k}(n)$ be the smallest prime greater than $n^k$, and let $d(n) = \dval$. Suppose that, for every constant $\delta > 0$, no uniform bounded-error quantum algorithm decides $k\OV$ in $O(n^{k/2 - \delta})$ time. Then, for every constant $\varepsilon > 0$, there is no uniform quantum algorithm $A$ that runs in $O(n^{k/2 - \varepsilon})$ time on every input and satisfies
\[
 \Pr_{x\sim\mathrm{Unif}(\mathbb F_{p(n)}^{knd(n)}),\,A}
 \bigl[A(x) = f\OV^k_{n,d(n),p(n)}(x)\bigr] \geq \frac34
\]
for all sufficiently large $n$. The probability is over the uniform input and the internal randomness of $A$.

\end{claim}

We overcome the prime computation overhead by subjecting it also to a quantum speedup. \cite{BHP01} showed that there unconditionally exists a prime in $[x, x+x^{0.525}]$ for all sufficiently large $x$. Therefore, we can use the well known quantum minimum finding algorithm by \cite{DH96} to search for a prime over this gap. This can easily be done in $\softO \left( n^{0.2625k} \right)$ time, well within our worst case lower bound. We formally state the proof below:

\begin{proof}

Suppose, towards a contradiction, that a uniform quantum algorithm $A$ computes $f\OV^k_{n,d,p_k}$ correctly on $3 / 4$ fraction of inputs in $O \left( n^{k/2-\epsilon} \right)$ time, for some $\epsilon > 0$ and for all large enough $n$. We give an algorithm $B'$ that computes $k\OV$ (for all large enough $n$) as follows:

\begin{algorithm}[h]
\caption{Algorithm $B'$}
\begin{algorithmic}[1]

\State Run the Dürr-Høyer minimum finding algorithm \citep{DH96} on the range $[n^k, n^k + n^{0.525k}]$ with the following objective function (oracle): Let $x = n^k$ and $L = \left \lceil x^{0.525} \right \rceil$, with a constant-factor enlargement if required by the precise prime-gap statement. Define, for $j \in \{\, x + 1, \ldots, x + L \,\}$,
\begin{equation*}
h(j) = 
\begin{cases}
j, & j \text{ is prime}, \\
x + L + 1, & j \text{ is composite}.
\end{cases}
\end{equation*}

\State $p \gets$ output of the Dürr-Høyer search \citep{DH96}

\State Run classic random self-reduction on $A$ to compute $\FOV^k_{n,d,p_k}$ on the worst case.

\State Let $c$ be the output of the amplified polynomial corrector on $(U_1, \ldots, U_k)$.

\State \Return $(c\neq 0)$.

\end{algorithmic}
\end{algorithm}

We can easily construct the primality testing circuit in $O(\polylog(n)) = \softO(1)$ \citep{AKS04} and therefore step 1 runs in $\softO \left( n^{0.2625k} \right)$. Step 3 runs in $\softO \left( n^{k/2-\epsilon} \right)$ time and therefore the total running time of algorithm $B'$ is 
$$T_{B'}(n) = \softO(n + n^{0.2625k} + n^{k/2 - \epsilon}) = \softO \left( n^{k/2-\epsilon} \right),$$ 
on the worst case for $k \geq 3$ and sufficiently small $\epsilon > 0$. The probability of success of step 1 and step 3 can be amplified accordingly so that the overall success probability is $> 2 / 3$. Therefore $B'$ decides $\kOV$\ efficiently on the worst case, which is a contradiction.

\end{proof}

\begin{remark}

The larger-prime construction uses an asymptotic prime-gap theorem. We do not derive an explicit numerical threshold $n_0$, concrete gate counts, or implementation benchmarks. Consequently, the present analysis does not determine which field choice is preferable
at practical input sizes. Explicit prime-gap estimates for other exponents are available \citep{DUD16, HUG23}; comparing these choices requires a separate analysis of thresholds, constants, and primality-testing costs.

\end{remark}

\section{Discussion and Open Problems}
\label{sec:discussion}

After looking at our results it would be natural to wonder why we are even concerned about uniform algorithms if we have shown hardness for non-uniform algorithms (which subsume uniform algorithms). The reason is that the underlying assumptions are different. Uniform hardness of $\FOV^k$ (in general) is based on $\SETH$ (or $\QSETH$) while non-uniform hardness of $\FOV^k$ is based on $\NUSETH$ (or $\NUQSETH$). While we believe $\NUSETH$ to be true, $\SETH$ is naturally a weaker assumption. Assuming $\SETH$ gives us the win-win situation since we do not expect advice while practically implementing efficient algorithms for natural fine-grained hard problems like \textit{Longest Common Subsequence} or \textit{Edit Distance}.

\subsection{Quantum Complexity of Exact Evaluation}

The classical complexity of $\FOV^k$ is characterized up to subpolynomial factors. Direct evaluation takes $\widetilde{O} \left( n^k \right)$ time, and the $k\OV$ conjecture implies an $\Omega \left( n^{k - o(1)} \right)$ average-case lower bound \citep{BRSV18}. In the quantum setting, our results establish an $\Omega \left( n^{k/2 - o(1)} \right)$ average-case lower bound for uniform algorithms under $\QSETH$ and the parameters specified in Section~\ref{sec:quniform}. The corresponding upper bound for
exact evaluation remains open.

The situation differs for the decision problem $k\OV$. Grover's algorithm \citep{Gro96} searches the $n^k$ candidate tuples in $\widetilde{O} \left( n^{k/2} \right)$ time when testing a tuple takes polylogarithmic time. This matches the conditional lower bound
established in Appendix~\ref{app:A}. However, this algorithm only determines whether an orthogonal tuple exists. For evaluating $\FOV^k$, our work does not provide a matching quantum algorithm. Thus, the classical $\widetilde{O} \left( n^k \right)$ evaluation algorithm remains the best known upper bound for $\FOV^k$.

This gap raises the following question.

\begin{quote}

\emph{Does exact evaluation of $\FOV^k$ admit a quantum algorithm with running time $O \left( n^{k/2+o(1)} \right)$, either in the worst-case or on uniformly random inputs? In particular, is the lower bound in Claim~\ref{claim:quniform} asymptotically tight?}

\end{quote}

An algorithm with this running time would match our lower bound up to subpolynomial factors. Conversely, a stronger lower bound would establish that exact evaluation retains computational difficulty beyond the quadratic speedup achieved by Grover's
algorithm \citep{Gro96} for the underlying decision problem.   

\subsection{Quantum Proof of Work}

A second natural direction, which we have not pursued here, is to come up with an efficient quantum analogue of the \textit{Proof of Work} construction of \cite{BRSV18}. This would not only require an efficient method of computing $\FOV^k$ but also need to show an efficient method by which the prover can prove the value of $\FOV^k$ to the verifier. While the classical protocol uses a modified version of the sum-check protocol to prove $\FOV^k$, the overhead of such a protocol will be too high for a quantum protocol where the amount of work expected is lower.

\bibliographystyle{apalike}

\bibliography{main}

\begin{thebibliography}{}

\bibitem[Aaronson et~al., 2020]{ACLWZ20}
Aaronson, S., Chia, N.-H., Lin, H.-H., Wang, C., and Zhang, R. (2020).
\newblock {On the Quantum Complexity of Closest Pair and Related Problems}.
\newblock In {\em 35th Computational Complexity Conference (CCC 2020)}, pages
  16:1--16:43.

\bibitem[Aggarwal and Kwan, 2025]{AK25}
Aggarwal, D. and Kwan, D. (2025).
\newblock {Quantum Worst-Case to Average-Case Reduction for Matrix-Vector
  Multiplication}.
\newblock arXiv:2510.15721.

\bibitem[Agrawal et~al., 2004]{AKS04}
Agrawal, M., Kayal, N., and Saxena, N. (2004).
\newblock Primes is in p.
\newblock {\em Annals of Mathematics}, 160(2):781--793.

\bibitem[Asadi et~al., 2024]{AGGSS24}
Asadi, V.~R., Golovnev, A., Gur, T., Shinkar, I., and Subramanian, S. (2024).
\newblock {Quantum Worst-Case to Average-Case Reductions for All Linear
  Problems}.
\newblock In {\em Proceedings of the 2024 Annual ACM-SIAM Symposium on Discrete
  Algorithms (SODA)}, pages 2535--2567.

\bibitem[Asadi et~al., 2026]{AHS26}
Asadi, V.~R., Hirahara, S., and Shimizu, N. (2026).
\newblock {Improved Quantum Random Self-Reduction for Linear Problems}.
\newblock arXiv:2609.39823.

\bibitem[Baker et~al., 2001]{BHP01}
Baker, R.~C., Harman, G., and Pintz, J. (2001).
\newblock The difference between consecutive primes, ii.
\newblock {\em Proceedings of the London Mathematical Society}, 83(3):532--562.

\bibitem[Ball et~al., 2024]{MJPAG24}
Ball, M., Garay, J., Hall, P., Kiayias, A., and Panagiotakos, G. (2024).
\newblock Towards permissionless consensus in the standard model via
  fine-grained complexity.
\newblock In {\em Advances in Cryptology – CRYPTO 2024: 44th Annual
  International Cryptology Conference}, pages 113--146.
\newblock Full version: Cryptology ePrint Archive, Report 2024/637,
  \url{https://eprint.iacr.org/2024/637}.

\bibitem[Ball et~al., 2017]{BRSV17}
Ball, M., Rosen, A., Sabin, M., and Vasudevan, P.~N. (2017).
\newblock Average-case fine-grained hardness.
\newblock In {\em Proceedings of the 49th Annual ACM SIGACT Symposium on Theory
  of Computing (STOC 2017)}, pages 483--496.
\newblock Full version: \url{https://eprint.iacr.org/2017/202}.

\bibitem[Ball et~al., 2018]{BRSV18}
Ball, M., Rosen, A., Sabin, M., and Vasudevan, P.~N. (2018).
\newblock Proofs of work from worst-case assumptions.
\newblock In {\em Advances in Cryptology -- CRYPTO 2018}, pages 789--819.

\bibitem[Cully-Hugill, 2023]{HUG23}
Cully-Hugill, M. (2023).
\newblock Primes between consecutive powers.
\newblock {\em Journal of Number Theory}, 247:100--117.

\bibitem[Dudek, 2016]{DUD16}
Dudek, A.~W. (2016).
\newblock {An explicit result for primes between cubes}.
\newblock {\em Functiones et Approximatio Commentarii Mathematici},
  55(2):177--197.

\bibitem[Durr and Hoyer, 1999]{DH96}
Durr, C. and Hoyer, P. (1999).
\newblock A quantum algorithm for finding the minimum.
\newblock arXiv:quant-ph/9607014v2.

\bibitem[Feigenbaum and Fortnow, 1993]{FF90}
Feigenbaum, J. and Fortnow, L. (1993).
\newblock Random-self-reducibility of complete sets.
\newblock {\em SIAM Journal on Computing}, 22(5):994--1005.

\bibitem[Gajulapalli et~al., 2026]{GKASS26}
Gajulapalli, K., Golovnev, A., King, S., and Saraogi, S. (2026).
\newblock Online orthogonal vectors revisited.
\newblock In {\em Proceedings of the 2026 Annual ACM-SIAM Symposium on Discrete
  Algorithms (SODA)}, pages 5641--5668.

\bibitem[Gemmell and Sudan, 1992]{GS92}
Gemmell, P. and Sudan, M. (1992).
\newblock Highly resilient correctors for polynomials.
\newblock {\em Information Processing Letters}, 43(4):169--174.

\bibitem[Grover, 1996]{Gro96}
Grover, L.~K. (1996).
\newblock A fast quantum mechanical algorithm for database search.
\newblock In {\em Proceedings of the 28th Annual ACM Symposium on Theory of
  Computing (STOC 1996)}, pages 212--219.

\bibitem[Hoeffding, 1963]{Hoeffding63}
Hoeffding, W. (1963).
\newblock Probability inequalities for sums of bounded random variables.
\newblock {\em Journal of the American Statistical Association},
  58(301):13--30.

\bibitem[Impagliazzo and Paturi, 2001]{Impagliazzo2001b}
Impagliazzo, R. and Paturi, R. (2001).
\newblock On the complexity of $k$-{SAT}.
\newblock {\em Journal of Computer and System Sciences}, 62(2):367--375.

\bibitem[Impagliazzo et~al., 2001]{IPZ01}
Impagliazzo, R., Paturi, R., and Zane, F. (2001).
\newblock Which problems have strongly exponential complexity?
\newblock {\em Journal of Computer and System Sciences}, 63(4):512--530.

\bibitem[Lagarias and Odlyzko, 1987]{LO87}
Lagarias, J. and Odlyzko, A. (1987).
\newblock Computing $\pi$(x): An analytic method.
\newblock {\em Journal of Algorithms}, 8(2):173--191.

\bibitem[Lipton, 1989]{Lip91}
Lipton, R.~J. (1989).
\newblock New directions in testing.
\newblock In {\em Distributed Computing and Cryptography: Proceedings of a
  DIMACS Workshop}, volume~2, pages 191--202.

\bibitem[Stephens-Davidowitz and Vaikuntanathan, 2019]{Stephens2019}
Stephens-Davidowitz, N. and Vaikuntanathan, V. (2019).
\newblock {SETH}-hardness of coding problems.
\newblock In {\em 2019 IEEE 60th Annual Symposium on Foundations of Computer
  Science (FOCS)}, pages 287--301.

\bibitem[Williams, 2005]{Wil05}
Williams, R. (2005).
\newblock A new algorithm for optimal 2-constraint satisfaction and its
  implications.
\newblock {\em Theoretical Computer Science}, 348(2):357--365.

\end{thebibliography}

\appendix
\section{Quantum Hardness of $k\OV$}
\label{app:A}

\begin{claim}[Claim for Theorem~\ref{thm:uqkov}]
\label{claim:uqkov}

Assuming $\QSETH$, for every fixed integer $k \geq 3$ and every constant $\varepsilon>0$, no uniform bounded-error quantum algorithm decides $k\OV$ on vectors of dimension $d(n) = \dval$ in $O(n^{k/2 - \varepsilon})$ time.

\end{claim}

\begin{proof}

We use the sparsification lemma \citep{IPZ01} to ensure the moderate dimension of all the $\kOV$\ instances.

\begin{lemma}[Sparsification Lemma \citep{IPZ01}]
\label{lem:sparsification}

For every constant $r \ge 2$ and every $\gamma > 0$ there is a constant $c = c(r, \gamma)$ and an algorithm that, given an $r$CNF formula $F$ on $N$ variables, outputs $r$CNF formulas $F_1,\ldots,F_t$ on the same variables such that

\begin{enumerate}

\item $t \le 2^{\gamma N}$,
    
\item $F$ is satisfiable if and only if at least one $F_s$ is satisfiable,
  
\item each $F_s$ has at most $c\cdot N$ clauses, and
  
\item the algorithm runs in time $2^{\gamma N} \poly(N)$.
    
\end{enumerate}
\end{lemma}

Fix $k\geq 3$ and suppose that a uniform quantum algorithm $Q$ decides $k\OV$ with error at most $1 / 3$ on every input of dimension $d(n) = \dval$, in time $O(n^{k/2 - \epsilon})$ for some constant
$\epsilon > 0$. Set
\[
 \beta = \min\{\epsilon,(k - 2)/4\},
 \qquad \gamma = \frac{\beta}{2k}.
\]
The same algorithm has running time $O(n^{k/2 - \beta})$. Let $r\geq 3$ be any fixed clause width and let $F$ be an $r$CNF formula on $N$ variables. We may add fewer than $k$ unused variables so that $k$ divides $N$; this changes the final exponential time bound by only a constant factor. Set
\[
 n = 2^{N/k},\qquad m = d(n) = \left\lceil(N/k)^2\right\rceil.
\]
Apply Lemma~\ref{lem:sparsification} with parameter $\gamma$. It produces $t \leq 2^{\gamma N}$ formulas $F_1, \ldots, F_t$ on these same variables, each with at most $cN$ clauses, where $c = c(r,\gamma)$ is independent of $N$. For all sufficiently large $N$, we have $cN \leq m$. The finitely many smaller lengths can be handled by exhaustive search without affecting the asymptotic bound. Pad each $F_s$ to exactly $m$ clauses using tautologies, such as $x_1 \lor \neg x_1$, so that satisfiability is preserved.

For each $s \in [t]$, write the clauses of its padded formula as $C_1, \ldots, C_m$. Partition the variables into $k$ blocks
\[
 X_i = \{\, x_{(i - 1) N/k + 1},\ldots,x_{iN/k} \,\},
 \qquad i\in[k].
\]
Each block has exactly $n$ partial assignments. We use one vector for each such assignment, obtaining $k$ indexed lists of $n$ vectors of dimension $m$.

Define $k$ maps,
\[
u_1, \ldots, u_k : \{\, 0, 1 \,\}^{N / k} \to \{\, 0, 1 \,\}^m.
\]
Each $u_i$ takes as input an arbitrary assignment of the variables in $X_i$ (say $\alpha$) and returns
\[
u_i(\alpha)_1, \ldots, u_i(\alpha)_m
\]
as follows:
\[
u_i(\alpha)_j =
\begin{cases}
0 & \text{if } C_j \text{ is satisfied by partial assignment } \alpha, \\
1 & \text{otherwise},
\end{cases}
\]
for all $i \in \{\, 1, \ldots, k \,\}$ and all $j \in \{\, 1, \ldots, m \,\}$.

\begin{lemma}
\label{passign}

For any $k$ partial assignments $\alpha_1, \ldots, \alpha_k$,
\[
\langle u_1(\alpha_1), \ldots, u_k(\alpha_k) \rangle \neq 0
\]
implies that there exists a clause $C_j$ which is not satisfied by any partial assignment. Moreover,
\[
\langle u_1(\alpha_1), \ldots, u_k(\alpha_k) \rangle = 0
\]
implies that every clause $C_j$ has at least one partial assignment satisfying it, and hence $\alpha_1, \ldots, \alpha_k$, where $\alpha_i$ is an assignment of the variables in $X_i$ for all $i \in \{\, 1, \ldots, k \,\}$, form a satisfying assignment for our $\SAT$ instance.
 
\end{lemma}

\begin{proof}
If
\[
\langle u_1(\alpha_1), \ldots, u_k(\alpha_k) \rangle \neq 0,
\]
then there exists an index $j$ such that
\[
u_i(\alpha_i)_j = 1 \quad \forall i \in \{\, 1, \ldots, k \,\}.
\]
By definition of $u_i$, this implies that clause $C_j$ is not satisfied by any of the partial assignments.

Conversely, if
\[
\langle u_1(\alpha_1), \ldots, u_k(\alpha_k) \rangle = 0,
\]
then for every $j$ there exists some $i$ such that
\[
u_i(\alpha_i)_j = 0.
\]
Hence every clause $C_j$ is satisfied by at least one partial assignment. Therefore, the combined assignment
$\alpha_1, \ldots, \alpha_k$ satisfies the entire $\SAT$ instance.

\end{proof}

By Lemma~\ref{passign}, the instance constructed from $F_s$ is a YES instance precisely when $F_s$ is satisfiable. If $t = 0$, return False immediately; otherwise assume $t \geq 1$. Because we may make $t$ bounded-error calls, a constant error bound for each call is insufficient. Amplify $Q$ by independent repetitions and majority vote to obtain $Q_{\mathrm{amp}}$ with error at most $1/(6t)$ on every input. An odd number $R = O(\log(6t)) = O(N)$ of repetitions suffices as follows: Fix an input, and let $X_i$ indicate that the $i$th independent execution of $Q$ returns an incorrect answer. Then $\mathbb{E}[X_i] \leq 1/3$. By the additive Chernoff--Hoeffding bound~\citep{Hoeffding63},
\[
\Pr\!\left[\text{majority vote is incorrect}\right]
\leq
\Pr\!\left[\sum_{i = 1}^{R}X_i \geq R/2\right]
\leq \exp(-R/18).
\]
Choosing $R$ to be the smallest odd integer satisfying $R\geq 18\ln(6t)$ therefore gives error at most $1/(6t)$ on every input. Since $t\leq 2^{\gamma N}$, we have $R=O(N)$. Each repetition starts with fresh work registers, and its output bit is measured before the majority is computed. The resulting SAT algorithm is as follows, with the finitely many small input lengths handled as above.

\begin{algorithm}[H]
\caption{Algorithm $A(F)$ with amplified $k\OV$ calls}
\begin{algorithmic}[1]
\State Set $n = 2^{N/k}$, $m = \left \lceil \log^2 n \right \rceil$, and $\gamma = \beta/(2k)$.
\State Apply Lemma~\ref{lem:sparsification} to obtain $F_1, \ldots, F_t$.
\If{$t = 0$}
    \State \Return False
\EndIf
\State Choose an odd $R = O(\log(6t))$ giving per-call error at most $1/(6t)$.
\For{$s = 1, \ldots, t$}
    \State Pad $F_s$ with tautologies to exactly $m$ clauses.
    \For{$i = 1, \ldots, k$}
        \For{each assignment $\alpha\in\{\, 0, 1 \,\}^{N/k}$ to $X_i$}
            \State Put $u_i(\alpha)$ into the indexed list $U_i$.
        \EndFor
    \EndFor
    \State Run $Q(U_1, \ldots, U_k)$ independently $R$ times.
    \If{a majority of the measured output bits are True}
        \State \Return True
    \EndIf
\EndFor
\State \Return False
\end{algorithmic}
\end{algorithm}

By a union bound, the probability that any amplified call is incorrect is at most $t / (6t) = 1/6$.
When all calls are correct, Lemmas~\ref{passign} and~\ref{lem:sparsification} imply that $A$ returns the correct answer for $F$. Thus $A$ decides $r\SAT$ with success probability at least $5 / 6$ on every input.

For each $F_s$, constructing the lists costs $O(knm) = 2^{N/k}\operatorname{poly}(N)$ time. The amplified call costs
\[
 O\!\left(Rn^{k/2 - \beta}\right)
 = 2^{(1/2 - \beta/k)N}\operatorname{poly}(N).
\]
Since $\beta\leq(k - 2) / 4$, we have $1/k \leq 1/2 - \beta/k$. Including sparsification and all $t$ instances gives
\begin{align*}
 T_A(N)
 &\leq 2^{\gamma N}
       \left(1 + 2^{N/k} + 2^{(1/2 - \beta/k)N}\right)
       \operatorname{poly}(N)\\
 & = 2^{(1/2 - \beta/(2k))N}\operatorname{poly}(N)\\
 & = O\!\left(2^{(1/2 - \beta/(4k))N}\right).
\end{align*}
The saving $\beta/(4k) > 0$ is independent of the fixed clause width $r$.
Writing $\zeta = \beta/(2k)$, the final bound is
$O(2^{(1 - \zeta)N/2})$ for every fixed $r$.
This contradicts $\QSETH$.
Hence no such uniform bounded-error quantum algorithm $Q$ exists.

\end{proof}

\begin{corollary}

Assuming $\NUQSETH$, for every fixed integer $k \geq 3$ and every constant $\varepsilon > 0$, no non-uniform bounded-error quantum algorithm decides $k\OV$ with $d(n) = \dval$ in $O(n^{k/2 - \varepsilon})$ time using $O(n^{k/2 - \varepsilon})$ classical bits of advice (Theorem~\ref{thm:nuqkov}).

\end{corollary}

Suppose such a non-uniform algorithm exists for some $\epsilon > 0$. Use the same $\beta$, $\gamma$, padding, and amplification as in Claim~\ref{claim:uqkov}. All sparsified instances have the same $n$ and the same dimension $m$, so their encodings have the same length. The advice string for the $k\OV$ algorithm therefore depends only on that length and is identical for every $F_s$ and every repetition. Supply this one string as the advice for the SAT algorithm. Classical advice can be reread or copied; neither the number of sparsified formulas nor amplification multiplies the required advice length. Writing $a_N$ for this advice string, its length satisfies
\[
 |a_N| = O(n^{k/2 - \epsilon})
      =O(2^{(1/2 - \beta/(4k))N}),
\]
because $\beta \leq \epsilon$. The running time is also $O(2^{(1/2 - \beta/(4k))N})$, and the success probability is at least $5/6$ on every input. As before, the same positive exponential saving applies for every fixed clause width $r$, contradicting $\NUQSETH$.

\end{document}